\documentclass{ws-procs961x669}            

\newcommand{\cH}{\mathcal{H}}
\newcommand{\rL}{\rho_\Lambda}
\newcommand{\CC}{\Lambda}
\newcommand{\xib}{\overline{\xi}}
\newcommand{\bk}{{\bf k}}
\newcommand{\rv}{\rho_{\rm vac}}
\newcommand{\nueff}{\nu_{\rm eff}}
\newcommand{\tnueff}{\tilde{\nu}_{\rm eff}}

\begin{document}

\title{Renormalized $\rv$ without $m^4$ terms}

\author{Cristian Moreno-Pulido\footnote{Speaker} and Joan Sol\`a Peracaula}

\address{Departament de F\'isica Qu\`antica i Astrof\'isica,
and
Institute of Cosmos Sciences,
Universitat de Barcelona,
Av. Diagonal 647, E-08028 Barcelona, Catalonia, Spain\\
E-mails: cristian.moreno@fqa.ub.edu, sola@fqa.ub.edu}
\begin{abstract}

The cosmological constant (CC) term, $\CC$, in Einstein's equations has been for about three decades a fundamental building block of the concordance or standard $\CC$CDM model of cosmology.  Although the latter is not free of fundamental problems,  it provides a good phenomenological description of the overall cosmological observations. However, an interesting improvement in such a phenomenological description and also a change in the theoretical status of the $\CC$-term occurs upon realizing that the vacuum energy is actually a ``running quantity" in quantum field theory in curved spacetime. Several  works have shown that this option can compete with the $\CC$CDM with a rigid $\CC$ term. The so-called, ``running vacuum models" (RVM) are characterized indeed by a vacuum energy density, $\rv$, which is evolving with time as a series of even powers of the Hubble parameter and its time derivatives. This form has been motivated by semi-qualitative renormalization group arguments in previous works. Here we  review a recent detailed computation by the authors of the renormalized energy-momentum tensor of a non-minimally coupled  scalar field with the help of adiabatic regularization procedure. The final result is noteworthy:   $\rv (H)$ takes the precise  structure of the RVM,  namely a constant term plus a dynamical component $\sim H^2$ (which should be detectable in the present universe)  including also higher order effects ${\cal O}(H^4)$ which can be of interest during the early stages of the cosmological evolution.  Besides, it is most remarkable that such renormalized form of the vacuum energy density does not carry dangerous terms proportional to $m^4$, the quartic powers of the masses of the fields, which are a well-known source of exceedingly large contributions to the vacuum energy density and are directly responsible for extreme fine tuning in the context of the cosmological constant problem.

\end{abstract}

\keywords{Cosmology, Dark Energy, Quantum Field Theory.}

\bodymatter

\section{Introduction}\label{Sec1:Intro}

Since the mid nineties, the cosmological constant (CC) term,  $\CC$,  in Einstein's equations has been a crucial ingredient of the `concordance' or standard  $\CC$CDM model of cosmology\,\cite{Peebles1984}. From the phenomenological point of view it was favored only as of the time $\CC$  became a physically measured quantity some twenty years ago\,\cite{SNIa}. Nowadays, precision cosmology\,\cite{PlanckCollab} is able to do better measurements of $\CC$, or more precisely of the associated parameter  $\Omega^0_{\rm vac}=\rho^0_{\rm vac}/\rho^0_{c}$, established by observations to be $\Omega^0_{\rm vac}\sim 0.7$. Here, the vacuum energy density (VED) is associated to $\Lambda$ and defined as $\rho^0_{\rm vac}=\CC /(8\pi G_N)$, where $G_N$ is Newton's constant and $\rho^0_{c}=3H_0^2/(8\pi G_N)$ is the critical density in our times. However, there is a lot of uncertainty with the CC with respect its nature and origin at a fundamental level. For instance it is commonly assumed that $\CC$ may be related with the Zero-Point Energy (ZPE) of the quantum matter fields and with the Higgs potential of the Standard Model (SM). However, if ones compares the observational value $\rv$ with the ZPE of a typical SM particle such as the electron we are left with a tiny value of $\rv^{\rm obs}/\rho_{\rm ZPE}\sim 10^{-34}$.  If compared with the ground state energy of the effective Higgs potential, $V_{\rm eff}$, we find $\rv^{\rm obs}/\langle V_{\rm eff}\rangle\sim 10^{-56}$. This issue is generically called the cosmological constant problem\,\cite{Weinberg89,Witten2000} and affects all forms of dark energy (DE)\,\cite{Sahni2000,PeeblesRatra2003,Padmanabhan2003,Copeland2006,DEBook}. Another aspect of this intriguing topic is the fact that, at the present time, $\rv^{\rm obs}$ and $\rho_{\rm CDM}$, the energy density associated to Cold Dark Matter (CDM),  are observed to be of the same order of magnitude, despite the fact that $\rho_{\rm CDM}$ is assumed to decrease with the expansion as $a^{-3}$, being $a$ the scale factor, whereas $\rv$ maintains constant. 

The astonishing theoretical problems, though, are not the only ones concerning the concordance model. Currently, some important measurements seems to be in tension with the $\CC$CDM, emphasizing the discrepant values of the Hubble parameter at the present time, $H_0$, obtained from independent measurements of the early and the local universe\,\cite{TensionsLCDM}. It is still unknown if these tensions are the result of systematic errors, but the chance that a deviation from the $\CC$CDM model could provide an explanation for such discrepancies\cite{Riess2019} is totally open.

As it has been shown in the literature, models mimicking a time-evolving  $\rho_{\rm vac}$, which are allowed by the cosmological principle, could help in alleviating the mentioned problems, see e.g.\,\cite{ApJL2015,RVMphenoOlder1,RVMpheno1,RVMpheno2,ApJL2019,Mehdi2019} and\cite{Vagnozzi,Valentino,OobaRatra,ParkRatra,MartinelliSalvatelli,Costa,LiZhang}. In this work, we review recent work\,\cite{Cristian2020} on the possibility of having dynamical vacuum energy density (VED), $\rv$, in the context of quantum field theory (QFT) in curved spacetime with a spatially flat Friedmann-Lema\^\i tre-Robertson-Walker (FLRW) metric. We want to focus on the dynamics associated to the running vacuum model (RVM)\,\cite{ShapSol1,Fossil2008,ShapSol2}; for a review, see \cite{JSPRev2013,JSPRev2015,AdriaPhD2017} and references therein. Related studies are e.g. \cite{Babic2005,Maggiore2011} and  \,\cite{KohriMatsui2017,Ferreiro2019}, but also others extending the subject to the context of supersymmetric theories \,\cite{Bilic2011,Bilic2012} and to supergravity\,\cite{SUGRA2015}. Recently, some works in the framework of the effective action of string theories\,\cite{BMS2020A,BMS2020B,Nick2021,NickPhiloTrans,NickInflationaryPhysics,NickGravitationalAnomalies} have been dealed with the topic. 

In the calculations reviewed here, the renormalization of the energy-momentum tensor is done using the adiabatic regularization procedure (ARP)\,\cite{BirrellDavies82,ParkerToms09,Fulling89,MukhanovWinitzki07}. The renormalization process is a WKB approximation of the field modes in an expanding FLRW background. Once the VED is renormalized, obtained upon inclusion of the renormalized value of $\rL$ (associated to $\CC$) at a given scale, it does not include the dangerous contributions depending on the fourth power of the particle masses ($\sim m^4$) and therefore it is free from huge induced corrections to the VED\,\cite{Cristian2020}. Additionally, we find a RVM-like form for the VED in the current times,  $c_0+\nu H^2$, with $c_0$ related to the usual constant term and $|\nu|\ll 1$.  This last parameter represents only a small (dynamical) correction to the constant term and, depending on the sign of $\nu$, it can mimic quintessence or phantom DE.

First, in Sec. \ref{sec:EMT} our framework is defined, consisting of a real scalar field non-minimally coupled to gravity, and we present the classical energy-momentum tensor (EMT). Later on, in Sections \ref{sec:AdiabaticVacuum} and \ref{sec:AREMT}, we present the quantum fluctuations in the adiabatic vacuum by means of a WKB expansion of the field modes in the  FLRW background and we discuss the adiabatic regularization of the EMT.
In Sec. \ref{sec:RenormZPE} we start with the renormalization of the EMT in the FLRW context through the ARP, which is then needed  in  Sec. \ref{sec:RenormalizedVED} to derive the precise result of $\rv$  from the renormalized ZPE up to terms of adiabatic order $4$. In our case that means up to  ${\cal O}(H^4)$. We also demonstrate that the values of the VED at different scales are related by an expression which does not contain quartic powers of the masses. To continue, in Sec. \ref{sec:RunningConnection}, we connect the computed VED in this work with the well-known running vacuum model (RVM), which originally was justified from renormalization group arguments in curved spacetime in previous works. A summary of the conclusions and discussion appear in Sec. \ref{sec:conclusions}. At the end, we define our conventions and collect some useful formulas in Appendix.

\section{Energy-Momentum Tensor for non-minimally coupled scalar field}\label{sec:EMT}

Einstein's equations read\,\footnote{Some useful geometrical quantities can be seen in the Appendix.}
\begin{equation}
R_{\mu \nu}-\frac{1}{2}R g_{\mu \nu}+\CC g_{\mu \nu}=8\pi G_N T_{\mu \nu}^{\rm matter}\,, \label{FieldEq}
\end{equation}
where $ T_{\mu \nu}^{\rm matter}$  is the EMT of matter.
They can also be written as
\begin{equation} \label{FieldEq2}
\frac{1}{8\pi G_N}G_{\mu \nu}+\rL g_{\mu \nu}=T_{\mu \nu}^{\rm matter}\,,
\end{equation}
where  $G_{\mu\nu}\equiv R_{\mu\nu}-\frac{1}{2}Rg_{\mu\nu}$ and  $\rL \equiv \CC /(8\pi G_N)$ is the VED associated to $\CC$. The associated contribution of $\CC$ to  $T_{\mu \nu}^\CC \equiv -\rL g_{\mu \nu}$. But, in general, there are more contributions to the total VED, associated to the quantum fluctuations of the fields, and also to their  classical ground state energy.  In this work, for practical reasons, we will assume that there is only one (matter) field in the form of a real scalar field, $\phi$, and we will denote by $T_{\mu \nu}^{\phi}$ the piece of the EMT associated to it.  Thus $T^{\rm tot}_{\mu\nu}=T_{\mu \nu}^\CC +T_{\mu \nu}^{\phi}$. We neglect other incoherent matter contributions such as dust and radiation because they can be added without altering the QFT considerations developed here.

A non-minimal coupling between the scalar field and gravity is assumed, without any classical potential for $\phi$ . The part of the action associated to $\phi$ is
\begin{equation}\label{eq:Sphi}
  S[\phi]=-\int d^4x \sqrt{-g}\left(\frac{1}{2}g^{\mu \nu}\partial_{\nu} \phi \partial_{\mu} \phi+\frac{1}{2}(m^2+\xi R)\phi^2 \right)\,,
\end{equation}
where $\xi$ is the non-minimal coupling between $\phi$ and gravity.  In the special case $\xi=1/6$, the massless ($m=0$)  action has conformal symmetry, i.e. symmetric under simultaneous rescalings of the $g_{\mu\nu}$ and $\phi$ with a local function $\alpha(x)$: $g_{\mu\nu}\to e^{2\alpha(x)}g_{\mu\nu}$ and  $\phi\to e^{-\alpha(x)}\phi$. However, we will keep $\xi$ general as it enables a richer phenomenology.

The Klein-Gordon (KG) equation is satisfied by the field $\phi$,
\begin{equation}
(\Box-m^2-\xi R)\phi=0.\label{KGequation}
\end{equation}
Here $\Box\phi=g^{\mu\nu}\nabla_\mu\nabla_\nu\phi=(-g)^{-1/2}\partial_\mu\left(\sqrt{-g}\, g^{\mu\nu}\partial_\nu\phi\right)$. On the other hand, the EMT takes this form :
\begin{equation}
\begin{split}
T_{\mu \nu}(\phi)=&-\frac{2}{\sqrt{-g}}\frac{\delta S_\phi}{\delta g^{\mu\nu}}= (1-2\xi) \partial_\mu \phi \partial_\nu\phi+\left(2\xi-\frac{1}{2} \right)g_{\mu \nu}\partial^\sigma \phi \partial_\sigma\phi\\
& -2\xi \phi \nabla_\mu \nabla_\nu \phi+2\xi g_{\mu \nu }\phi \Box \phi +\xi G_{\mu \nu}\phi^2-\frac{1}{2}m^2 g_{\mu \nu} \phi^2.
\end{split} \label{EMTScalarField}
\end{equation}
In this work we consider a spatially flat FLRW metric in the conformal frame. Being $\eta$ the conformal time we have  $ds^2=a^2(\eta)\eta_{\mu\nu}dx^\mu dx^\nu$, with $\eta_{\mu\nu}={\rm diag} (-1, +1, +1, +1)$ the Minkowski  metric.
Differentiation with respect the conformal time is denoted by $\left(\right)^\prime\equiv d\left(\right)/d\eta$. It is convinient to define $\cH(\eta)\equiv a^\prime /a$ and, since $dt=a d\eta$, it is related with the usual Hubble rate $H(t)=\dot{a}/a$ as $\mathcal{H}(\eta)=a  H(t)$. Despite the fact we will do intermediate calculations in conformal time, at the end it will be useful to express the VED in terms of the usual Hubble rate $H(t)$, because it is the easiest way to compare with other RVM results in the literature.

 \section{Quantum fluctuations and WKB ansatz}\label{sec:AdiabaticVacuum}

Now, we can take into account the quantum fluctuations of the field $\phi$ as an expansion around the background field (or classical mean field), $\phi_b$:
\begin{equation}
\phi(\eta,x)=\phi_b(\eta)+\delta \phi (\eta,x). \label{ExpansionField}
\end{equation}
The vacuum expectation value (VEV) of the field is $\langle 0 | \phi (\eta, x) | 0\rangle=\phi_b (\eta)$, whereas we assume that the VEV of the fluctuations vanishes:  $\langle 0 | \delta\phi | 0\rangle =0$. The vacuum state to which we are referring to is the so-called adiabatic vacuum, we will make some comments about this below.

The corresponding EMT decomposes itself as $\langle T_{\mu \nu}^\phi \rangle=\langle T_{\mu \nu}^{\phi_b} \rangle+\langle T_{\mu \nu}^{\delta_\phi}\rangle$, where
$\langle T_{\mu \nu}^{\phi_{b}} \rangle =T_{\mu \nu}^{\phi_{b}} $
is the  contribution  from the background part and  $\langle T_{\mu \nu}^{\delta\phi}\rangle$ is related with the quantum fluctuations. In particular, the  $\langle T_{00}^{\delta\phi}\rangle$ is associated to the ZPE density of the scalar field in the FLRW background. Thus, the total vacuum contribution to the EMT reads
\begin{equation}
\langle T_{\mu \nu}^{\rm vac} \rangle= T_{\mu \nu}^\CC +\langle T_{\mu \nu}^{\delta \phi}\rangle=-\rL g_{\mu \nu}+\langle T_{\mu \nu}^{\delta \phi}\rangle\,.\label{EMTvacuum}
\end{equation}
This means that the total vacuum EMT receives contributions from the cosmological constant term and from the quantum fluctuations. A renormalized version of this equation will lead us to a renormalized VED, as we will see later.

The KG equation (\ref{KGequation}) is satisfied independently by the classical field and the quantum part (\ref{ExpansionField}). Let us concentrate on the fluctuation $\delta\phi$.  Its Fourier decomposition in frequency modes $h_k(\eta)$ is
\begin{equation}
\delta \phi(\eta,{\bf x})=\frac{1}{(2\pi)^{3/2}a}\int d^3{k} \left[ A_\bk e^{i{\bf k\cdot x}} h_k(\eta)+A_\bk^\dagger e^{-i{\bf k\cdot x}} h_k^*(\eta) \right]\,. \label{FourierModes}
\end{equation}
$A_\bk$ and  $A_\bk^\dagger $ are the (time-independent) annihilation and creation operators, with commutation relations
\begin{equation}
[A_\bk, {A_{\bk'}}^\dagger]=\delta({\bf k}-{\bf k'}), \qquad [A_\bk,A_{\bk'}]=0. \label{CommutationRelation}
\end{equation}
Introducing the Fourier expansion (\ref{FourierModes}) in $\left(\Box-m^2-\xi R\right)\delta\phi=0$
we find that the frequency modes of the fluctuations satisfy
\begin{equation}
h_k^{\prime \prime}+ \Omega_k^2 h_k=0\,, \ \ \ \ \ \ \ \ \ \ \Omega_k^2(\eta) \equiv \omega_k^2(m)+a^2\, (\xi-1/6)R\,, \label{KGModes}
\end{equation}
with  $\omega_k^2(m)\equiv k^2+a^2 m^2$.
As we can see, $h_k$ depends uniquely on the modulus $k\equiv|\bk|$ of the momentum. Since $\Omega_k(\eta)$ is a nontrivial function of the conformal time, there is no closed form of the solution of (\ref{KGModes}). Instead of looking to an analytical solution, we will generate an approximate solution from  a recursive method based on the phase integral ansatz
\begin{equation}
h_k(\eta)\sim\frac{1}{\sqrt{W_k(\eta)}}\exp\left(i\int^\eta W_k(\tilde{\eta})d\tilde{\eta} \right)\,. \label{WKBSolution}
\end{equation}
Therefore, the function $W_k(\eta)$ satisfies the following differential equation:
\begin{equation}
W_k^2=\Omega_k^2 -\frac{1}{2}\frac{W_k^{\prime \prime}}{W_k}+\frac{3}{4}\left( \frac{W_k^\prime}{W_k}\right)^2\,. \label{WKBIteration}
\end{equation}
This non-linear differential equation can be solved using the WKB approximation. The solution is valid for large $k$ (i.e. for short wave lengths) and the function  $\Omega_k$ is slowly varying for weak fields. The notion of vacuum we work with it is called the adiabatic vacuum\,\cite{Bunch1980,ParkerFulling1974,BunchParker1979}, and can be defined as the quantum state annihilated by all the operators $A_k$ of the Fourier expansion of the scalar field, see \,\cite{BirrellDavies82,ParkerToms09,Fulling89,MukhanovWinitzki07} for details. The physical interpretation of the modes (\ref{KGModes}) with frequencies dependending on time, must be understood in terms of field observables rather than in particle language. Therefore, for a more physical interpretation of the vacuum effects of the expanding background, we must compute the  renormalized EMT in the FLRW spacetime, but first we need to regularize it.

\section{Adiabatic Regularization of the EMT}\label{sec:AREMT}

The adiabatic (slowly varying) $\Omega_k$, is susceptible to be computed using Eq.\,(\ref{WKBIteration}) to generate an (asymptotic) series solution. In this context, the series is obtained through the  adiabatic regularization procedure (ARP)\footnote{This method was introduced for minimally coupled massive scalar fields in\,\cite{ParkerFulling1974,BunchParker1979}. In \,\cite{Bunch1980}, it is generalized for other couplings. A review can be found in the classic books \cite{BirrellDavies82,ParkerToms09}. The ARP has been used in the context of QFT in curved backgrounds\,\cite{Ferreiro2019,KohriMatsui2017} and also it has been extended to other fields, such as the spin one-half fields in\,\cite{Landete,Barbero}.} and it is organized in what we call adiabatic orders. First, the quantites considered of adiabatic order 0 are: $k^2$ and $a$. Of adiabatic order 1 are: $a^\prime$ and $\cH$. Then $a^{\prime \prime},a^{\prime 2},\cH^\prime$ and $\cH^2$ are quantities of adiabatic order 2. We can sum up by saying that each extra derivative in conformal time increases the adiabatic order one unit. As a consequence, the  ``effective frequency'' $W_k$ can be written as an asymptotic expansion:
\begin{equation}\label{WKB}
W_k=\omega_k^{(0)}+\omega_k^{(2)}+\omega_k^{(4)}+\dots,
\end{equation}
where each $\omega_k^{(j)}$ is an adiabatic correction of order $j$.  This leads to an expansion of the mode function $h_k$ in even order  adiabatic terms. This is justified by arguments of general covariance, since only terms of even adiabatic order (an even number of time derivatives) are allowed in the field equations.

\subsection{Introducing the renormalization parameter}\label{sect:RelatingScales}

We start by defining the 0\textit{th} order terms
\begin{equation}\label{omegak0}
\omega_k^{(0)} \equiv\omega_k= \sqrt{k^2+a^2 M^2} .
\end{equation}
In this approach the WKB expansion is performed off-shell, at an arbitrary mass scale $M$ replacing the scalar field mass $m$ in \eqref{omegak0}. In consequence, ARP can be formulated in such a way that we can relate the adiabatically renormalized theory at two different scales\,\cite{Ferreiro2019}. If $M = m$  we obtain the renormalized theory on-shell. In the computation of the EMT, the parameter $\Delta^2\equiv m^2-M^2$ will appear in the correction terms and it has to be considered of adiabatic order 2 since it appears in the WKB expansion together with other terms of the same adiabatic order\,\cite{Ferreiro2019}. If $\Delta = 0$, then $M = m$ and corresponds to the usual on-shell ARP, see \,\cite{BirrellDavies82,ParkerToms09}.
With the help of this formalism one can explore the evolution of the VED throughout the cosmological history as it has been done for the first time in \,\cite{Cristian2020}. As we shall see, it will be convenient to consider $m$ to be of the order of magnitude as the masses of Grand Unified Theory (GUT) fields in order to explore the dynamics of the VED in the low energy domain $M^2\ll m^2$ (corresponding to the late universe). For simplicity, we model here all particles in terms of real scalar fields as in  \,\cite{Cristian2020}. For a generalization to fermion fields, see \,\cite{Samira2021}.

\subsection{Regularized ZPE}\label{eq:RegZPE}

Our starting point is the initial solution $W_k\approx \omega_k^{(0)}$ indicated in Eq. \eqref{omegak0}.  For  $a=1$, it corresponds to the standard Minkowski space modes. Since $a=a(\eta)$ we have to find a better approximation.  We use the initial solution $\omega_k^{(0)}$ in  \eqref{WKBIteration} and expand the RHS in powers of  $\omega_k^{-1}$, then collect the new terms up to adiabatic order 2 to find $\omega_k^{(2)}$. Now, we repeat the process with $W_k\approx \omega_k^{(0)}+\omega_k^{(2)}$ on the RHS of the same equation, expand again  in  $\omega_k^{-1}$, collect contributions of adiabatic order 4... After some steps, the expansion seems to be organized in powers of  $\omega_k^{-1}\sim 1/k$ (i.e. a short wavelength expansion). The UV divergent terms of the ARP are precisely the first lowest powers of $\omega_k^{-1}$, which are present in the first adiabatic orders of the expansion. Higher adiabatic orders come later in the iteration, and  represent finite contributions \cite{Bunch1980,ParkerFulling1974,BunchParker1979} since they decay quickly at large $k$ and the associated integrals are manifestly convergent. In our case, the divergent terms of the EMT are present up to $4{\it th}$ adiabatic order so that, at least, we have to compute all the terms up to this order (more detail below in Eq.\,\eqref{EMTFluctuations}). After renormalization, we will obtain a finite expression for the EMT and then compute the vacuum energy density. But first, we need to consider the EMT associated to the fluctuations, using Eq.\,\eqref{EMTScalarField} and (\ref{ExpansionField}). For the $00$-component,
\begin{equation}
\begin{split}\label{EMTInTermsOfDeltaPhi}
\langle T_{00}^{\delta \phi} \rangle &= \Big\langle \frac{1}{2}\left(\delta\phi^{\prime}\right)^2  + \left(\frac{1}{2}-2\xi\right)\sum_i\partial_i \delta \phi\partial_i \delta \phi\\
&+6\xi\cH\delta \phi \delta \phi^\prime-2\xi\delta \phi \nabla^2 \delta\phi+3\xi\cH^2\delta \phi^2+\frac{a^2m^2}{2}\left(\delta \phi \right)^2 \Big\rangle ,\\
\end{split}
\end{equation}
where $\nabla^2 \equiv \sum_{i=1}^3 \partial^2_i$ and $\left(\delta\phi^{\prime}\right)^2\equiv\left(\delta\partial_0\phi\right)^2= \left(\partial_0\delta\phi\right)^2$.  We may now substitute the Fourier expansion of $\delta\phi$, as given in \eqref{FourierModes},  into Eq.\,\eqref{EMTInTermsOfDeltaPhi} and apply the commutation relations \eqref{CommutationRelation}.
After symmetrizing the operator field product $\delta\phi \delta\phi^\prime$ with respect to the  annihilation and creation operators, we end up with
\begin{equation}
\begin{split}
\langle T_{00}^{\delta \phi} \rangle & =\frac{1}{8\pi^2 a^2}\int dk k^2 \left[ 2\omega_k+\frac{a^4M^4 \cH^2}{4\omega_k^5}-\frac{a^4 M^4}{16 \omega_k^7}(2\cH^{\prime\prime}\cH-\cH^{\prime 2}+8 \cH^\prime \cH^2+4\cH^4)\right.\\
&+\frac{7a^6 M^6}{8 \omega_k^9}(\cH^\prime \cH^2+2\cH^4) -\frac{105 a^8 M^8 \cH^4}{64 \omega_k^{11}}\\
&+\overline{\xi}\left(-\frac{6\cH^2}{\omega_k}-\frac{6 a^2 M^2\cH^2}{\omega_k^3}+\frac{a^2 M^2}{2\omega_k^5}(6\cH^{\prime \prime}\cH-3\cH^{\prime 2}+12\cH^\prime \cH^2)\right. \\
& \left. -\frac{a^4 M^4}{8\omega_k^7}(120 \cH^\prime \cH^2 +210 \cH^4)+\frac{105a^6 M^6 \cH^4}{4\omega_k^9}\right)\\
&+\left. \xib^2\left(-\frac{1}{4\omega_k^3}(72\cH^{\prime\prime}\cH-36\cH^{\prime 2}-108\cH^4)+\frac{54a^2M^2}{\omega_k^5}(\cH^\prime \cH^2+\cH^4) \right)
\right]\\
&+\frac{1}{8\pi^2 a^2} \int dk k^2 \left[  \frac{a^2\Delta^2}{\omega_k} -\frac{a^4 \Delta^4}{4\omega_k^3}+\frac{a^4 \cH^2 M^2 \Delta^2}{2\omega_k^5}-\frac{5}{8}\frac{a^6\cH^2 M^4\Delta^2}{\omega_k^7} \right.\\
& \left. + \xib \left(-\frac{3a^2\Delta^2 \cH^2}{\omega_k^3}+\frac{9a^4 M^2 \Delta^2 \cH^2}{\omega_k^5}\right)\right]+\dots, \label{EMTFluctuations}
\end{split}
\end{equation}
where we have defined $\xib\equiv \xi-1/6$ and we have integrated $ \int\frac{d^3k}{(2\pi)^3}(...)$ over solid angles and expressed the final integration in terms of $k=|\bk|$. Let us note the presence of the $\Delta$-dependent terms in the last two rows, which contribute at 2{\it th }($\Delta^2$) and 4{\it th } ($\Delta^4$) adiabatic order.

\section{Renormalization of the ZPE in curved spacetime}\label{sec:RenormZPE}

The ZPE part of the EMT, as given by  Eq.\,\eqref{EMTFluctuations} can be split into two parts as follows:
\begin{equation}
\langle T_{00}^{\delta \phi}\rangle (M)= \langle T_{00}^{\delta \phi}\rangle_{Div}(M)+\langle T_{00}^{\delta \phi}\rangle_{Non-Div}(M), \label{DecompositionEMT}
\end{equation}
where
\begin{equation}
\begin{split}
&\langle T_{00}^{\delta \phi}\rangle_{Div}(M)=\frac{1}{8\pi^2 a^2}\int dk k^2 \Bigg[ 2\omega_k +\frac{a^2 \Delta^2}{\omega_k}-\frac{a^4\Delta^4}{4\omega_k^3}\\
& -6\xib \cH^2\left(\frac{1}{\omega_k}+\frac{a^2 M^2}{\omega_k^3}+\frac{a^2 \Delta^2}{2\omega_k^3}\right) -9\frac{\xib^2}{\omega_k^3}(2\cH^{\prime\prime}\cH-\cH^{\prime 2}-3\cH^4) \Bigg],
\end{split} \label{DivergentPart}
\end{equation}
which containts the powers  $1/\omega_k^n$ up to $n=3$, manifestly UV-divergent.

On the other hand, the non-divergent part of (\ref{DecompositionEMT}) involves the integrals with powers of  $1/\omega_k$ higher than $3$ which are perfectly finite. Computing these convergent integrals (see Eq.\,(\ref{DRFormula}) in Appendix), the result reads
\begin{equation}\label{Non-DivergentPart}
\begin{split}
&\langle T_{00}^{\delta \phi}\rangle_{Non-Div}(M) =\frac{m^2 \cH^2}{96\pi^2}-\frac{1}{960\pi^2 a^2}\left( 2\cH^{\prime \prime}\cH-\cH^{\prime 2}-2\cH^4\right)+ \xib \frac{3\Delta^2 \cH^2}{8\pi^2}\\
&+\frac{1}{16\pi^2 a^2}\xib \left(2\cH^{\prime \prime}\cH-\cH^{\prime 2}-3\cH^4\right)+\frac{9}{4\pi^2 a^2}\xib^2 (\cH^\prime \cH^2 +\cH^4)+\dots\\
\end{split}
\end{equation}
where the dots in the last expression correspond to higher adiabatic orders.
Let us now focus on the divergent part of the ZPE, Eq.\,\eqref{DivergentPart}. The adiabatic series is an asymptotic series representation of  Eq.\,\eqref{EMTInTermsOfDeltaPhi}. Thefore, such series is not convergent but it provides an approximation, which is obtained once the series is cut at a particular order\,\cite{Ferreiro2019}. As a consequence there is some arbitrariness in the way of choosing the leading adiabatic order, which we can use in our favour.  There is, nonetheless, some previous steps to do before obtaining a meaningful result. First, we are going to set the arbitrary scale at the mass of the scalar field. That is, $M=m$ and hence $\Delta=0$ (cf. Sec.\ref{sect:RelatingScales}). The divergent part (\ref{DivergentPart}) is reduced in this case to
\begin{equation}
\begin{split}
&\langle T_{00}^{\delta \phi}\rangle_{Div}(m)=\frac{1}{8\pi^2 a^2}\int dk k^2 \Bigg[ 2\omega_k(m) -\xib  6\cH^2\left(\frac{1}{\omega_k(m)}+\frac{a^2m^2}{{\omega_k^3(m)}}\right) \\
& -\xib^2\frac{9}{{\omega_k^3(m)}}(2\cH^{\prime\prime}\cH-\cH^{\prime 2}-3\cH^4) \Bigg]\,.
\end{split} \label{DivergentPartClassic}
\end{equation}
What we are going to do next, in order to renormalize the ZPE and the EMT, is to subtract the terms that appear up to $4{\it th}$ adiabatic order at the arbitrary mass scale $M$.  This procedure is enough to cancel the divergent terms through the ARP, as it is discussed in the literature\,\cite{BirrellDavies82,ParkerToms09,Fulling89}.

\subsection{Renormalized ZPE for generic M}\label{sec:ZPEoffshell}

As said before, our proposal for the renormalized  ZPE in curved spacetime at the scale $M$ is \,\cite{Cristian2020}:
\begin{eqnarray}\label{EMTRenormalized}
\langle T_{00}^{\delta \phi}\rangle_{\rm Ren}(M)&=&\langle T_{00}^{\delta \phi}\rangle(m)-\langle T_{00}^{\delta \phi}\rangle^{(0-4)}(M)\nonumber\\
&=&\langle T_{00}^{\delta \phi}\rangle_{Div}(m)-\langle T_{00}^{\delta \phi}\rangle_{Div}(M)-\xib\frac{3\Delta^2 \cH^2}{8\pi^2}+\dots,
\end{eqnarray}
here $(0-4)$ means that the expansion is up to fourth adiabatic order and the dots in (\ref{EMTRenormalized}) denote perfectly finite terms of higher adiabatic order.  Using now Eq.\,\eqref{DivergentPartClassic}, we arrive to
\begin{equation}
\begin{split}
&\langle T_{00}^{\delta \phi}\rangle_{\rm Ren}(M)=\frac{1}{8\pi^2 a^2}\int dk k^2 \left[ 2 \omega_k (m)-\frac{a^2 \Delta^2}{\omega_k (M)}+\frac{a^4 \Delta^4}{{4\omega^3_k (M)}}-2 \omega_k (M)\right]\\
&+\xib \frac{6\cH^2}{8\pi^2 a^2}\Bigg\{ \int dk k^2 \left[\frac{1}{\omega_k (M)}+\frac{a^2 M^2}{{\omega^3_k (M)}}+\frac{a^2 \Delta^2}{2{\omega^3_k (M)}}-\frac{1}{\omega_k(m)}-\frac{a^2 m^2}{{\omega^3_k (m)}} \right]-\frac{a^2\Delta^2}{2}\Bigg\}\\
&-\xib^2 \frac{9\left(2 \cH^{\prime \prime}\cH-\cH^{\prime 2}-3 \cH^{4}\right)}{8\pi^2 a^2}\int dk k^2 \left[ \frac{1}{{\omega^3_k (m)}}-\frac{1}{{\omega^3_k (M)}}\right]+\dots
\end{split} \label{Renormalized}
\end{equation}
In this equation we have introduced new notation in order to distinguish between the off-shell energy mode $\omega_k(M)\equiv \sqrt{k^2+a^2 M^2}$  (formerly denoted just as $\omega_k$) and the on-shell one $\omega_k(m)\equiv \sqrt{k^2+a^2 m^2}$.
The involved computations of these convergent integrals are not so straightforward but, after some algebra and the help of Mathematica\cite{Mathematica}, we find
\begin{equation}
\begin{split}
&\langle T_{00}^{\delta \phi}\rangle_{\rm Ren}(M)=\frac{a^2}{128\pi^2 }\left(-M^4+4m^2M^2-3m^4+2m^4 \ln \frac{m^2}{M^2}\right)\\
&-\xib\frac{3 \cH^2 }{16 \pi^2 }\left(m^2-M^2-m^2\ln \frac{m^2}{M^2} \right)+\xib^2 \frac{9\left(2  \cH^{\prime \prime} \cH- \cH^{\prime 2}- 3  \cH^{4}\right)}{16\pi^2 a^2}\ln \frac{m^2}{M^2}+\dots
\end{split} \label{RenormalizedExplicit2}
\end{equation}
The obtained expression vanishes for $M=m$ as expected from Eq.\,(\ref{EMTRenormalized}).
However, let us remind the reader that this only happens because we have computed the on-shell value $\langle T_{\mu\nu}^{\delta \phi}\rangle_{\rm Ren}(m)$ up to adiabatic order $4$ in Eq.\,(\ref{RenormalizedExplicit2}). Nevertheless, $\langle T_{\mu\nu}^{\delta \phi}\rangle_{\rm Ren}(m)$ can be computed up to an arbitrary, but finite, adiabatic order. Beyond $4{\it th}$ order one always obtains subleading and perfectly finite corrections. We expect that these effects become suppressed when the physical mass $m$  of the quantum field is large, so that they satisfy the Appelquist-Carazzone decoupling theorem\,\cite{AppelquistCarazzone75}.

So far, we have obtained a renormalized ZPE in curved spacetime (\ref{RenormalizedExplicit2}) which, despite the fact it is finite, it includes the quartic powers of the masses corresponding to undesired contributions to the vacuum energy.

\section{Renormalized vacuum energy density}\label{sec:RenormalizedVED}

The renormalization in the context of QFT in curved spacetime implies the consideration of the higher derivative (HD) terms in the classical effective action\,\cite{BirrellDavies82}, beyond the usual Einstein-Hilbert (EH) term with a cosmological constant, $\CC$. In particular, in the FLRW background in four dimensions, the only surviving HD term is $H_{\mu \nu}^{(1)}$, as suggested by the geometric structure of \,(\ref{RenormalizedExplicit2}).  $H_{\mu \nu}^{(1)}$ is  obtained by functionally differentiating  $R^2$ with respect the metric (see Appendix). So, the full action contains the EH+HD terms and also the matter part which, in our simplified case, only consists of a scalar field $\phi$ non-minimally coupled to gravity, Eq.\,(\ref{eq:Sphi}). As usual, the modified Einstein's equations are obtained through the variation of the new action with respect to $g_{\mu\nu}$, extending \eqref{FieldEq2} as follows:
\begin{equation}\label{eq:MEEs}
\frac{1}{8\pi G_N(M)}G_{\mu \nu}+\rL(M) g_{\mu \nu}+a_1(M) H_{\mu \nu}^{(1)}= T_{\mu \nu}^{\phi_b} +\langle T_{\mu \nu}^{\delta \phi} \rangle_{\rm Ren}(M)\,,
\end{equation}
where the dependence in $M$ indicates we are considering renormalized quantities. In particular, the classical part of the matter EMT does not depend on it.
Our goal is to relate the  theory at different renormalization points\footnote{We do not pretend to compute the renormalized couplings from first principles, in particular the VED. Renormalization program allows us to compare the theory at different renormalization points but, at the end, an input from the experiment is needed to predict its value at another scale.}, and with this purpose in mind let us subtract Einstein's equations as written in \eqref{eq:MEEs} at two different scales, $M$ and $M_0$. We  find
\begin{equation}
\langle T_{\mu \nu}^{\delta \phi}\rangle_{\rm Ren}(M)- \langle T_{\mu \nu}^{\delta \phi}\rangle_{\rm Ren}(M_0)=f_{G_N^{-1}}  G_{\mu \nu}+f_{\rL} g_{\mu \nu}+f_{a_1} H^{(1)}_{\mu \nu},  	\label{EinsteinDifferentScale}
\end{equation}
here $f_X(m,M,M_0)\equiv X(M)- X(M_0)$ for the various couplings involved $X=G_N^{-1}, \rL, a_1$.
Since we know the expression of the renormalized EMT within the ARP, namely  Eq\, (\ref{RenormalizedExplicit2}),  we can obtain the renormalization shift of the couplings  $G_N^{-1}$, $\rL$  and $a_1$ in \eqref{EinsteinDifferentScale} between the two scales $M$ and $M_0$.  This is possible by taking the value of the components $G_{00}$ and $H_{00}^{(1)}$ (from Appendix) and comparing with (\ref{RenormalizedExplicit2}).  Notice that the first term of (\ref{RenormalizedExplicit2}) , in the \textit{r.h.s}, is of 0{\it th} adiabatic order and is associated to $f_{\rL}(m,M,M_0)$ and, therefore, to the running  of $\rL$.
By doing this comparision we find that the functions are
\begin{equation} \label{SubtractionOfCoupling}
\begin{split}
&f_{G_N^{-1}}(m,M,M_0)=\frac{\xib}{16\pi^2}\left[M^2 - M_0^{2} -m^2\ln \frac{M^{2}}{M_0^2}\right], \\
&f_{\rL}(m,M,M_0) =\frac{1}{128\pi^2}\left(M^4-M_0^{4}-4m^2(M^2-M_0^{2})+2m^4\ln  \frac{M^{2}}{M_0^2}\right),\\
& f_{a_1}(M,M_0)= -\frac{\xib^2}{32\pi^2} \ln \frac{M^2}{M_0^{2}}.\\
\end{split}
\end{equation}

\subsection{Running Vacuum Energy density. Absence of $\sim m^4$ terms.}\label{sec:TotalVED}

At this point we can come back to the definition of vacuum energy \eqref{EMTvacuum} which was used in this work.  Considering $\langle T_{\mu\nu}^{{\rm vac}}\rangle=p_{\rm vac}g_{\mu \nu}+\left(p_{\rm vac}+\rv\right)u_\mu u_\nu$, with $u_\mu=(a,0,0,0)$, and equating with  Eq.\,\eqref{EMTvacuum},
and taking the $00$-component of the equality (keeping also in mind that $g_{00}=-a^2(\eta)$ in the conformal frame), we obtain
\begin{equation}\label{eq:Totalrhovac}
\rv(M)=\rL(M)+\frac{\langle T_{00}^{\delta \phi}\rangle_{\rm Ren}(M )}{a^2}.
\end{equation}
We write the explicit dependence in the renormalization point $M$ since we are talking about the renormalized quantity at that scale.
Equation (\ref{eq:Totalrhovac}) means that the total VED at an arbitrary  scale $M$ is not only receiving contributions from the renormalized cosmological term but also from the quantum fluctuations of fields (i.e. from the renormalized ZPE of the scalar field, in our simplified model).  Subtracting the renormalized result at two scales, $M$ and $M_0$, and using \eqref{EinsteinDifferentScale}, we find:
\begin{equation}
\begin{split}
&\rv(M)-\rv(M_0)=\rL (M)-\rL (M_0)+\frac{\langle T_{00}^{\delta \phi}\rangle_{\rm Ren}(M)- \langle T_{00}^{\delta \phi}\rangle_{\rm Ren}(M_0)}{a^2}\\
&=f_{\rL}(m,M,M_0)+\frac{f_{G_N^{-1}}(m,M,M_0)  G_{00}+f_{\rL}(m,M,M_0) g_{00}+f_{a_1}(M,M_0) H^{(1)}_{00}}{a^2}\\
&=\frac{f_{G_N^{-1}}(m,M,M_0) }{a^2}  G_{00}+\frac{f_{a_1}(M,M_0)}{a^2} H^{(1)}_{00}\\
&=\frac{3\cH^2}{a^2}\,f_{G_N^{-1}}(m,M,M_0)-\frac{18}{a^4}\left(\cH^{\prime 2}-2\cH^{\prime \prime}\cH+3 \cH^4 \right)\,f_{a_1}(M,M_0)\,,\phantom{\,\,\,aaaaaaaaa}.\label{RenormalizedVEa}
\end{split}
\end{equation}
As we see the term $f_{\rL}(m,M,M_0)$ has disappeared. Finally, from (\ref{SubtractionOfCoupling}) we obtain\,\cite{Cristian2020}
\begin{equation}
\begin{split}
\rv(M)
=&\rv(M_0)+\frac{3}{16\pi^2}\xib H^2\left[M^2 - M_0^{2}-m^2\ln \frac{M^{2}}{M_0^2}\right]\\
&+\frac{9}{16\pi^2} \xib^2 \left(\dot{H}^2 - 2 H\ddot{H} - 6 H^2 \dot{H} \right)\ln \frac{M^2}{M_0^2}\,.\label{CosmicTimeRenormalizedVE}
\end{split}
\end{equation}
In addition, we have used Eq.\,(\ref{eq:H100}) from Appendix to write the final result in terms of the Hubble function in cosmic time ($H=\cH/a$). The  result \eqref{CosmicTimeRenormalizedVE} is  the value of the VED at a particular energy scale $M$, related with the value of the VED at another renormalization scale $M_0$, i.e. it  expresses the `running' of the VED.  Notice that if $\xi=1/6$ ($\xib = 0$), then the VED is independent of the renormalization scale. However, this is not likely to happen since one also has to consider the contribution from fermions and vector boson fields. It is also important to notice that the running  is slow for small $H$, as it depends as ${\cal O}(H^2)$ times a mass scale squared and on ${\cal O}(H^4)$ contributions. Finally, we remark that there is no contributions of quartic mass scales.

\section{Running vacuum connection}\label{sec:RunningConnection}

As explained in a footnote in previous sections, the result we were seeking (represented now by Eq.\,\eqref{CosmicTimeRenormalizedVE}) was not the calculated value of the VED at a given scale, e.g. it says nothing on the actual value of $\rv(M_0)$ and hence it has nothing to do with the cosmological constant problem mentioned in the Introduction. In other words, we have not provided the calculation of the value itself of the vacuum energy at particular point of the cosmic history.  This result can nevertheless be useful to study the `running' of the VED from one scale to another. We can rephrase this by saying that if $\rv$ is known at some scale $M_0$, Eq.\,\eqref{CosmicTimeRenormalizedVE} can  be used to obtain the value of $\rv$ at another scale $M$. Such connection was suggested long ago from renormalization group arguments in curved spacetime\,\cite{ShapSol1,Fossil2008,ShapSol2} --  see  \cite{JSPRev2013,JSPRev2015}  and references therein for a review of the running vacuum model (RVM). What is more, it can even set a framework for the possible variation of the so-called fundamental constants of nature\,\cite{FritzschSola} with the evolution of the universe.
In the next subsection we are going to suggest a possible interpretation of  Eq.\,\eqref{CosmicTimeRenormalizedVE} in the context of the RVM. For this purpose, let us assume that we set one arbitrary scale to the renormalized VED at some Grand Unified Theory (GUT) scale  (i.e., set $M_0=M_X$, where $M_X\sim 10^{16}$ GeV). This value is associated also with the inflationary scale.  It is natural to assume that fundamental parameters, such as e.g.  $\rv$,   are  defined at that primeval scale, related with the very beginning in the cosmological history. Additionally, a large scale such as the GUT scale insures that all particle masses can be active degrees of freedom to some extent.

\subsection{RVM in the currect universe}

Equation \eqref{CosmicTimeRenormalizedVE} can, in principle, be used to explore the value of the VED throughout the cosmological history. However, for the study of the very early universe (in particular, inflation) other contributions can appear, which will not be addressed here since in the present work we are just going to study the current universe dynamics in order to compare with the late time RVM phenomenology.  For related studies in different contexts, cf. \,\cite{Basilakos2013,SolaGV2015,Sola2015,Yu2020}. See also \,\cite{Nick2021,NickPhiloTrans,NickInflationaryPhysics,NickGravitationalAnomalies} for a stringy version of the RVM with implications on the mechanism of inflation.

Consider $\rv(M_X)$, the value of the  VED at $M_0=M_X$. The value of this energy density is unknown, but our aim is to relate it with the current value of the VED, $\rv^0\equiv \rv (H_0)$,  by fixing the numerical value of the second scale at $M=H_0$, where  $H_0$  today's value of the Hubble function-- it can be considered an estimation for the energy scale associated to the the present FLRW universe.  This association is commonly made in the aforementioned references on the RVM\,\cite{JSPRev2013}. So, \eqref{CosmicTimeRenormalizedVE} applied to the present universe is
\begin{equation}\label{eq:rlMX1}
\rv^0\approx \rv(M_X)-\frac{3}{16\pi^2}\xib H_0^2\left[M_X^2+{m^2}\ln \frac{H_0^{2}}{M_X^2}\right].\\
\end{equation}
Here we ignore all terms of order  ${\cal O}(H^4)$ (including $\dot{H}^2, H\ddot{H}$ and  $H^2 \dot{H}$)  for the present universe ($H=H_0$). We can use this relation to find the value of $\rv(M_X)$, and the result is
\begin{equation}\label{eq:rlMX2}
\rv(M_X)=\rv^0-\frac{3\nueff}{8\pi}\,H_0^2\,M_P^2\,,
\end{equation}
where we have defined $\nueff$, the `running  parameter' for the VED:
\begin{equation}\label{eq:nueff}
\nueff=-\frac{\xib}{2\pi}\frac{M_X^2}{M_P^2}\left(1+\frac{m^2}{M_X^2}\ln \frac{H_0^{2}}{M_X^2}\right)\,.
\end{equation}
Notice that for $\xi=1/6$ (or $\xib=0$), it vanishes and there is no running, as mentioned previously. In the case $\xi=0$ (or $\xib=-1/6$) and $m^2/M_X^2\ll1$ we obtain $\nueff\simeq\frac{1}{12\pi}\,\frac{M_X^2}{M_P^2}\ll1$. More in general, the structure  obtained for  $\nueff$  is very similar to that one obtained previously in other works with the RVM approach, see \cite{JSPRev2013}. In these contexts, the parameter is related with the one-loop $\beta$-function for the renormalization group equation of $\rv$.  Although, in our case, there is and additional logarithmic contribution $\ln {H_0^{2}}/{M_X^2}$, coming from the direct QFT calculation developed here. At the end of the day it does not make significant differences in practice since $\nueff$ is a constant parameter which is fitted directly to cosmological data as an effective coefficient. It is natural to expect a small $\nueff$, i.e. $|\nueff|\ll1$, because of the ratio $M_X^2/M_P^2\sim 10^{-6}$, but notice that $\nueff$ depends on $\xi$ and also on contributions from other fields (fermions and bosons) and their multiplicities. Heavy fields of the order of the GUT scale can contribute in a significant way ($m\sim M_X$). As mentioned, $\nueff$ needs to be fitted the RVM to tbe cosmological data, and this has been done in detail e.g. in \cite{RVMpheno1,RVMpheno2}, the results indicate that $\nueff\sim 10^{-3}$.

Now, we can us Eq.\,\eqref{CosmicTimeRenormalizedVE} and \,\eqref{eq:rlMX2} to study the late universe. By ignoring ${\cal O}(H^4)$, we may estimate the current VED by  taking  $M = H$ around the present times:
\begin{equation}\label{eq:RVM1}
\rv(H)=\rv^0-\frac{3\nueff}{8\pi}\,H_0^2\,M_P^2+\frac{3\nueff (H)}{8\pi}\,H^2\,M_P^2\,,
\end{equation}
where
\begin{equation}\label{eq:nueffH}
\nueff (H)=-\frac{\xib}{2\pi}\frac{M_X^2}{M_P^2}\left(1+\frac{m^2}{M_X^2}\ln \frac{H^{2}}{M_X^2}\right)\,.
\end{equation}
Notice that the last expression is not constant as a difference from \eqref{eq:nueff}. However, since the dynamical behaviour of $\nueff(H)$ is logarithmic, and considering a value of $H$ close to $H_0$, we can approximate $\nueff(H)$ by \,\eqref{eq:nueff}.  Then, equation (\ref{eq:RVM1}) may be written as\,\cite{Samira2021}
\begin{equation}\label{eq:RVM2}
\rv(H)\simeq \rv^0+\frac{3\nueff}{8\pi}\,(H^2-H_0^2)\,M_P^2=\rv^0+\frac{3\nueff}{8\pi G_N}\,(H^2-H_0^2)\,.
\end{equation}
This is exactly the canonical form of the  RVM formula\,\cite{JSPRev2013}.  Such approximation holds fine  even if we study the CMB epoch, since the difference between $\nueff (H)$ from $\nueff$ can be estimated to be less than $8\%$ (for $m\simeq M_X$) or much less when $m\ll M_X$.  This is the typical parametrization that has been used in the literature about the RVM \cite{RVMpheno1,RVMpheno2}. As a matter of fact, this parametrization can be promoted to a more general form, by considering also a term proportional to $\dot{H}$ in the running equation for the VED. In the following subsection we are going to give grounds for the appearence of these kind of factors, not yet present in
\,\eqref{eq:RVM2}.

\subsection{More geometric structures for vacuum in curved spacetime}

Using the definition of the EMT associated to vacuum given in \,\eqref{EMTvacuum} we have derived the expression of the vacuum energy density in equation \,\eqref{eq:Totalrhovac}. However, we expect that the definition \,\eqref{EMTvacuum} can be generalized as follows:
\begin{equation}\label{eq:GeneralVDE1}
 \left\langle T_{\mu\nu}^{\rm vac} \right\rangle=T_{\mu\nu}^\CC +\left\langle T_{\mu\nu}^{\delta\phi} \right\rangle + \alpha_1 R g_{\mu\nu} +\alpha_2 R_{\mu\nu}+\mathcal{O}(R^2).
\end{equation}
$\mathcal{O}(R^2)$ represents tensors of adiabatic order 4, that is $R^2$, $R_{\mu\nu}R^{{\mu\nu}},\dots$  and $\alpha_i$ are parameters of dimension $+2$ in natural units. This new form for the vacuum EMT can be justified from the phenomenological point of view, but we also remind the reader that in a more realistic picture, contributions from other fields (bosons and fermions) are also expected, and general covariance leads to a generic form as represented in \,\eqref{eq:GeneralVDE1}. Generation of new terms is also possible by considering string-inspired mechanisms as it has been shown recently in \,\cite{BMS2020A,BMS2020B,Nick2021,NickPhiloTrans}.

Let us work with the more general form for the vacuum EMT, \eqref{eq:GeneralVDE1}.  It is useful to redefine the coefficients of that expression as $\alpha_i (M)\equiv \frac{\lambda_i}{16\pi^2} M^2$.  We can check in Appendix the expression of $R$ and  $R_{00}$ in flat FLRW spacetime, we obtain after a straightforward calculation:
\begin{equation}\label{eq:RVMgeneralized}
\rv(H)\simeq \rv^0+\frac{3}{8\pi G_N}\,\left(\tnueff\,H^2+ \bar{\nu}\,\dot{H}\right)\,,
\end{equation}
where
\begin{equation}\label{eq:nuefftilde}
\tnueff=\frac{1}{2\pi}\,\frac{M_X^2}{M_P^2}\,\left[\left(\frac{1}{6}-\xi\right)\left(1+\frac{m^2}{M_X^2}\ln \frac{H_0^{2}}{M_X^2}\right)+4\lambda_1+\lambda_2\right]\,
\end{equation}
and
\begin{equation}\label{eq:nubar}
\bar{\nu}=\frac{2\lambda_1+\lambda_2}{2\pi}\,\frac{M_X^2}{M_P^2}\,.
\end{equation}
As done before, the logarithmic part of $\tnueff(H)$ can be neglected and we may approximate $\tnueff(H_0)\simeq \tnueff$. The parameter $\bar{\nu}$ is treated as constant here, but we can not discard also a possible running of $\lambda_i$, acquiring a dependence in $M$. Although, if this were the case, it is reasonable to expect a logarithmic evolution of the renormalization effects. The above formula is a generalization of Eq.\,\eqref{eq:RVM2} incorporating the additional coefficient $\bar{\nu}$, accompanying $\sim\dot{H}$. 
Finally, let us comment that several generalized forms of the  RVM  beyond the usual one containing the $H^2 $ term has been studied before in the literature, see for instance \cite{ApJL2015,RVMphenoOlder1}.

\section{Discussion and conclusions}\label{sec:conclusions}

In this presentation we have investigated the possible dynamics of vacuum in the context of QFT in  FLRW spacetime. 
Specifically,  we have revisited the calculation of the renormalized energy-momentum tensor (EMT) of a real quantum scalar field non-minimally coupled to the FLRW background\cite{Cristian2020}.  The approach is based on adiabatic regularization and  renormalization of the EMT, starting from the WKB approximation of the field modes in curved spacetime. The renormalized EMT is defined as the difference of its on-shell value and its value at an arbitrary renormalization point $M$ up to 4{\it th} adiabatic order. It is sufficient with this subtraction since the divergent terms are present up to 4{\it th} adiabatic order. The renormalized EMT is perfectly finite and acquires a dependence on $M$. We can use this fact to compare the renormalized result at different times of the history of the universe characterized by different energy scales. At this point we can identify the VED from the renormalized  EMT. Such VED depends on $\rL(M)$ but also in the ZPE part involving the quantum fluctuations of the scalar field. An interesting implication is that the combination of both quantities is free from terms  $\sim m^4$, well-known sources of large contributions to the VED.

This QFT calculation leads us to a renormalized VED which has the usual form of the running vacuum models (RVM's)\,\cite{JSPRev2013}, in which $\rv=\rv(H)$ consists of an additive constant plus a series of powers of $H$ (the Hubble rate) and its time derivatives.  In previous works, the RVM  was motivated from general renormalization group arguments in QFT in curved spacetime (cf. \cite{JSPRev2013} and references therein). But, in this work, we have reviewed proof that the RVM form of the VED for the current universe emerges also from this calculation of QFT in the FLRW spacetime involving adiabatic regularization. It is found that the powers of $H$ (and its time derivatives) in the EMT carry an even number of time derivatives of the scale factor, which is mandatory if we impose the general covariance of the action.  In particular, the lowest order dynamical component of the VED is  $\sim\nu\,H^2$. The dimensionless coefficient $\nu$ is expected to be small  ($|\nu|\ll1$). However, at the end, the model has to be fitted with the help of a complet set of cosmological data.  This lower order term  $\sim\,H^2$ has the main role in the study of the dynamics of the vacuum in the late universe, whereas  the higher order terms may play a major role in the very early universe, for example to describe inflation (despite the fact that we have not exploited this idea in this work).
In previous works, the RVM has been succesfully confronted to a large number of cosmological data and the effective parameter $\nu$ has been fitted. The encountered results say that $\nu$ is positive and around $\sim 10^{-3}$, see\cite{ApJL2015,RVMphenoOlder1,RVMpheno1,RVMpheno2} . To conclude, let us remark that our QFT calculation has some limitations, for instance it has been simplified by the use of just one single field in the form of a real quantum scalar field. For this reason, further investigations have to be done in order to generalize these results for multiple fields involving scalar, fermionic and vectorial degrees of freedom. However, we do not expect major changes beyond computational details and the main results presented here are likely to be maintained\cite{Samira2021}.
 
 \vspace{-0.1cm}
\section*{Acknowledgements}
This work is based on the invited talk presented in the CM3 parallel session: ``Status of the $H_0$ and $\sigma_8$ Tensions: Theoretical Models and Model-Independent Constraints'' of the MG16 Marcel Grossmann virtual Conference, July 5-10 2021. We would like to thank the organizers of MG16 as well as the chairpersons of the session  for the excellence of the event and for the invitation. 

Work partially funded by  projects  PID2019-105614GB-C21 and FPA2016-76005-C2-1-P (MINECO, Spain), 2017-SGR-929 (Generalitat de Catalunya) and CEX2019-000918-M (ICCUB).  CMP  is also partially supported by the fellowship 2019 FI$_{-}$B 00351  (Generalitat de Catalunya).   JSP acknowledges participation in the COST Association Action CA18108 ``{\it Quantum Gravity Phenomenology in the Multimessenger Approach (QG-MM)}''.

\section*{Appendix : Conventions, geometrical quantities and useful formulas}
Through this work we have used natural units. This means that the gravitational constant is written as $G_N=1/M_P$, being $M_P$ is the Planck mass and  $\hbar=c=1$. 

Conventions regarding the geometrical quantities can be summarized as follows: the metric has  the signature $(-, +,+,+ )$; the Riemann tensor is
$R^\lambda_{\,\,\,\,\mu \nu \sigma} = \partial_\nu \, \Gamma^\lambda_{\,\,\mu\sigma} + \Gamma^\rho_{\,\, \mu\sigma} \, \Gamma^\lambda_{\,\, \rho\nu} - (\nu \leftrightarrow \sigma)$; the Ricci tensor is $R_{\mu\nu} = R^\lambda_{\,\,\,\,\mu \lambda \nu}$; and finally, Ricci scalar reads  $R = g^{\mu\nu} R_{\mu\nu}$. These elections correspond to $(+, +, +)$ in the Misner-Thorne-Wheeler\,\cite{MTW} conventions.

On the other hand, the Einstein tensor is defined as  $G_{\mu\nu}=R_{\mu\nu}-\frac12\,R g_{\mu\nu}$ and the field equations are $G_{\mu\nu}+\CC g_{\mu\nu}=8\pi G_N\, T_{\mu\nu}$. The (non-vanishing) Christoffel symbols, with a line element in terms of the conformal time $ds^2=a^2(\eta)\eta_{\mu\nu}dx^\mu dx^\nu$ and $\eta_{\mu\nu}={\rm diag} (-1, +1, +1, +1)$, are:
\begin{equation}
\Gamma_{00}^{0}=\cH,\qquad \Gamma_{ij}^0=\cH\delta_{ij}, \qquad \Gamma_{j0}^i=\cH\delta_j^i\,.
\end{equation}
The Ricci scalar and the 00-components of the curvature tensors are:
\begin{equation}\label{eq:R}
R={6}\frac{a^{\prime\prime}}{a^3}=\frac{6}{a^2}\,(\cH^\prime+\cH^2)=6\,\left(\frac{\dot{a}^2}{a^2}+\frac{\ddot{a}}{a}\right)=6(2H^2+\dot{H})
\end{equation}
and
\begin{equation}\label{eq:R00G00}
 \ \ \ R_{00}=-3\cH^\prime=-3a^2(H^2+\dot{H})\,,\qquad G_{00}=3\cH^2=3a^2H^2\,.
\end{equation}
Primes are indicating differentiation with respect to conformal time, meanwhile we use dots to represent derivatives with respect cosmic time.
The process of renormalization introduces the need for the  higher order curvature tensor $H_{\mu\nu}^{(1)}$, obtained by the variation of $R^2$ with respect the metric in the higher derivative vacuum action:
\begin{equation}\label{eq:H1munu}
  H_{\mu\nu}^{(1)}=\frac{1}{\sqrt{-g}}  \frac{\delta}{\delta g^{\mu\nu}}  \int d^4x  \sqrt{-g} R^2=-2\nabla_\mu\nabla_\nu R + 2 g_{\mu\nu} \Box R - \frac12 g_{\mu\nu} R^2 +2 R R_{\mu\nu}\,.
\end{equation}
In particular, the $00$-component is related with the vacuum energy density and in the conformally flat metric is
\begin{equation}\label{eq:H100}
H_{00}^{(1)}=\frac{-18}{a^2}\left(\cH^{\prime 2}-2\cH^{\prime \prime}\cH+3 \cH^4 \right)= -18 a^2\left(\dot{H}^2-2H\ddot{H}-6H^2\dot{H}\right)\,.
\end{equation}
\subsection*{Computing integrals with powers of $1/\omega_k$}
The following expression is useful to compute all the finite ($n>N$) and UV-divergent ($n\leq N$)  integrals involved in the EMT computation:
\begin{equation}\label{DRFormula}
I(n,Q)\equiv \int \frac{d^N k}{(2\pi)^N}\frac{1}{(k^2+Q^2)^{n/2}}=\frac{1}{(4\pi)^{N/2}}\frac{\Gamma\left(\frac{n-N}{2}\right)}{\Gamma\left( \frac{n}{2}\right)}\left({Q^2}\right)^{\frac{N-n}{2}},\\
\end{equation}
where $k\equiv|\bk|$ and $Q$ is an arbitrary scale. We have corrected typos in this formula as compared to that of Appendix B of   \cite{Cristian2020}, with no consequences in the calculation.

\bibliographystyle{ws-procs961x669}

\end{document}